\documentclass[aps,amsfonts,superscriptaddress, nofootinbib, floatfix, 10pt,twocolumn,prd]{revtex4-1} 
\usepackage{graphicx}
\usepackage{amsmath}
\usepackage[latin1]{inputenc}
\usepackage{amsbsy}
\usepackage{soul}
\usepackage{color}
\usepackage{epsfig}
\usepackage{graphicx}
\usepackage{amsmath}
\usepackage{amssymb}
\usepackage{bbm}
\usepackage{amsbsy}
\usepackage{slashed}

\usepackage{xcolor}

\usepackage[normalem]{ulem}  

\def\be{\begin{equation}}

\def\ee{\end{equation}}
\def\bea{\begin{eqnarray}}
\def\eea{\end{eqnarray}}

\begin{document}


\title{Magnetic renormalons in a scalar self interacting  $\lambda \phi^{4}$ theory}

\author{M. Correa}
\email{mcorrea52@uc.cl}
\affiliation{Instituto de F\'isica, Pontificia Universidad Cat\'olica de Chile, Casilla 306, San\-tia\-go 22, Chile}
\author{M.~Loewe}
\email{mloewe@fis.puc.cl}
\affiliation{Instituto de F\'isica, Pontificia Universidad Cat\'olica de Chile, Casilla 306, San\-tia\-go 22, Chile}
\affiliation{Centre for Theoretical and Mathematical Physics and Department of Physics,
University of Cape Town, Rondebosch 7700, South Africa}
\affiliation{Centro Cient\'ifico Tecnol\'ogico de Valpara\'iso-CCTVAL, Universidad T\'ecnica Federico Santa Mar\'ia, Casilla 110-V, Vapara\'iso, Chile} 
\author{D. Valenzuela}
\email{devalenz@gmail.com}
\affiliation{Instituto de F\'isica, Pontificia Universidad Cat\'olica de Chile, Casilla 306, San\-tia\-go 22, Chile}
\author{R. Zamora}
\email{rzamorajofre@gmail.com}
\affiliation{Instituto de Ciencias B\'asicas, Universidad Diego Portales, Casilla 298-V, Santiago, Chile} 
\affiliation{Centro de Investigaci\'on y Desarrollo de Ciencias Aeroespaciales (CIDCA), Fuerza A\'erea de Chile, Santiago 8020744, Chile}



\begin{abstract}

We present an analysis  about the influence of an external magnetic field on renormalons in a self interacting theory $\lambda \phi ^{4}$. In the weak magnetic field region, using an appropriate expansion for the Schwinger propagator's, we find renormalons as poles on the real axis of the Borel plane, whose position do not depend on the magnetic field, but where the residues acquire in fact a magnetic dependence. In the strong magnetic limit, working in the lowest Landau level approximation (LLLA), these new poles are not longer present. We compare the magnetic scenario with previous results in the literature concerning thermal effects on renormalons in this theory.

\end{abstract}

\maketitle


\section{Introduction}

Since the paper by Dyson \cite{Dyson} on the convergence of perturbative series in quantum electrodynamics (QED), a seminal article based exclusively on physical arguments, we have learned that power series expansions in Quantum Field Theory (QFT), in general, are divergent objects. For high orders  of perturbation, the divergence grows like $n!$, where n is the order of expansion, and this is due, essentially, to the multiplicity of diagrams that contribute to a certain Green function, or to a physical process, at such expansion orders. A usual procedure to improve the convergence relies on the Borel transform \cite{Altarelli,Rivasseau,Khanna}. However, in some cases, even the Borel transformed series  are divergent, spoiling  the meaning of the whole procedure. The new singularities responsible for this divergent behavior are the renormalons. For a review, see \cite{Beneke}. Recently there has been a renewed interest in the subject by considering one loop renormalization group equations in multi-field  theories \cite{Vasquez} or by considering finite temperature mass correction in the $\lambda \phi ^{4}$ theory, reanalyzing the temperature dependence of poles and their residues \cite{Cavalcanti}. There are other sources of divergences, as instantons in quantum chromodynamics (QCD) \cite{Gross,Shafer}, for example. However we will not refer here to such objects that can be handled using semi-classical methods. 

\bigskip
In peripheral heavy ion collisions, huge magnetic fields appear \cite{warringa}. In fact, the biggest fields existing in nature. The interaction between the produced pions in those collisions may be strongly affected by the magnetic field. In this article we analyze, in the frame of a self interacting scalar  $\lambda \phi^{4}$ theory, the influence of the magnetic field on the position of the U.V. renormalons (the only relevant in $\lambda\phi^4$ theory) and their residues in the Borel plane. For this purpose we will use the weak  field expansion for a bosonic Schwinger propagator \cite{Ayalaetal}. We present in detail the different analytical techniques we have used for our calculations, making also some comments on the strong magnetic case. The article is organized as follows. In Sec. \ref{magneticpropagator} we present the  bosonic propagator, in the presence of an external magnetic field, to be used in our work. In Sec. \ref{Boreltransform} we revisit  and explain the method of the Borel transform. In Sec. \ref{secRenVacio} we compute the renormalons for the $\lambda \phi ^{ 4}$ theory in the vacuum. In Sec. \ref{magneticrenormalons} we calculate the magnetic effect on the renormalons. Finally, we summarize and present the conclusions of our analysis in Sec. \ref{conclusions}.



\section{Magnetic renormalon-type correction to the propagator} \label{magneticpropagator}

Magnetic corrections will be handled through Schwinger's bosonic propagator given by \cite{Schwinger}
\begin{equation} \label{propfase}
D^B(x',x'')=\phi(x',x'')\int \frac{d^4k}{(2\pi)^4} e^{-ik\cdot(x'-x'')}D^B(k) \text{,}
\end{equation}
where

\begin{eqnarray}\label{prop1}
iD^B(k)&=&\int_0^\infty \frac{ds}{\cos(eBs)} \nonumber \\
&\times&\exp \left( is \left[ k_\parallel^2-k_\perp^2\frac{\tan (eBs)}{eBs}-m^2+i\epsilon \right] \right) \text{.}
\end{eqnarray}

\noindent The 4-momentum has been decomposed into parallel and perpendicular components respect to the magnetic field direction. By considering a constant magnetic field, whose direction defines the z-axis, we can write

\begin{equation}
(a \cdot b)_\parallel = a^0b^0 -a^3b^3 \text{,} \qquad (a \cdot b)_\perp = a^1b^1+a^2b^2 \text{,}
\end{equation}

\noindent for two arbitrary four vectors $a_{\mu}$ and $b_{\mu}$. We have also

\begin{equation}
a^2 = a_\parallel ^2 - a_\perp ^2 \text{.}
\end{equation}\\

\noindent Notice that the phase factor in Eq. (\ref{propfase}), given by

\begin{equation}\label{fase}
\phi(x',x'')=\exp \left(  ie \int_{x''}^{x'} dx_\mu A^\mu (x)  \right) \text{,}
\end{equation}

\noindent which is independent of the path, can be ignored for two-point functions, diagonal in configuration space. Since we will work with expressions at thee one loop, which are diagonal in the configuration space, this factor will disappear. Therefore, we shall work in the momentum representation.

In this way using $eBs \rightarrow s$ we find

\begin{eqnarray} \label{propmin}
iD^B(k)&=&\frac{1}{eB}\int_0^\infty \frac{ds}{\cos(s)} \nonumber \\
&&\times \exp \left( i(s/eB) \left[ k_\parallel^2-k_\perp^2\frac{\tan (s)}{s}-m^2+i\epsilon \right] \right) \text{.} \nonumber \\
\end{eqnarray}\\

This propagator can be expressed as a sum over Landau levels \cite{Ayalaetal}

\begin{equation}\label{proplandau}
iD^B(k)=2i \sum_{l=0}^\infty \frac{(-1)^l L_l \left( \frac{2k_\perp^2}{eB} \right) e^{-k_\perp^2/eB}}{k_\parallel^2-(2l+1)eB - m^2 +i\epsilon} \text{,}
\end{equation}
where $L_l$ are the Laguerre polynomials. By considering the previous expression in the region where $eB \ll m^{2}$ it is possible to show  that 

\begin{eqnarray}\label{propmagneticodebil}
iD^B(k)&&\xrightarrow{eB\rightarrow 0} \frac{i}{k_\parallel^2 - k_\perp^2 -m^2} - \frac{i(eB)^2}{(k_\parallel^2 - k_\perp^2 -m^2)^3} \nonumber \\
&-& \frac{2i(eB)^2k_\perp^2}{(k_\parallel^2 - k_\perp^2 -m^2)^4} \text{.}
\end{eqnarray}\\

which is the desired weak field expansion for our calculation.



\section{The Borel Transform} \label{Boreltransform}

We will briefly revise the Borel transform method, which is a tool designed to make sense to series potentially divergent series \cite{Altarelli}.\\
\\
For a divergent perturbative expansion 

\begin{equation}
D[a]=\sum_{n=1}^\infty D_na^n \text{.}
\end{equation}

The Borel transform $B[b]$ of the series $D[a]$ is defined through 

\begin{equation}
B[b]=\sum_{n=0}^\infty D_{n+1}\frac{b^n}{n!} \text{.}
\end{equation}

The inverse transform is, 

\begin{equation}
D[a]=\int_0^\infty db \hspace{2mm} e^{-b/a} B[b] \text{.}
\end{equation}\\

We need the last integral to be convergent, in order to make sense to the series, and $B[b]$ free form singularities in the range of integration. If this conditions are fulfilled we say that the original series $D[a]$ is Borel summable.\\

It is easy to check that all convergent series are also Borel summable. For the case of divergent series this is not necessary the case. If we find poles in the $0-\infty $ range of integration of the previous equation, the series is no longer Borel summable. It is possible to make sense to this integral in such cases, through a prescription for integration path in the complex $[b]$-plane, avoiding the pole. This will be, however, a prescription-dependent result.

A classical reference about divergent series is the book by Hardy, \cite{hardy}.

It is known that perturbative expansions in quantum field theory are not Borel summable. There are two sources for the appearance of singularities in the Borel plane: renormalons and instantons. Here we do not want to comment about the latter possibility. In QCD, renormalons have been a matter of debate since these objects affect our understanding of the gluon condensate \cite{Beneke}.

\section{Renormalons in the vacuum}\label{secRenVacio}
 In the $\lambda \phi^{4}$ theory the renormalon type diagrams that produce poles in the Borel plane correspond to a correction of the two-point green function

\vspace{0.5cm}
\begin{figure}[h] 
\begin{center} 
\includegraphics[width=7cm]{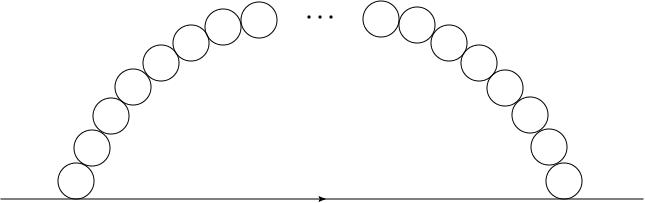}
\caption{Renormalon-type contribution to the two point function.}
\label{fig:diagrenormalon}
\end{center}
\end{figure}

First we revise the calculation of this diagram with the insertion of k bubbles, summing then over $k$ and studying the behavior of its transform in the Borel plane.\\

We will denote by $R_k(p)$ the diagram of order $k$ shown in Fig. {\ref{fig:diagrenormalon}}, where  $p$ is the four-momentum  entering and leaving the diagram and  $q$ is the four momentum that goes around the chain of bubbles

\begin{equation}\label{Rk}
R_k(p)=\int \frac{d^4q}{(2 \pi)^4}\frac{i}{(p+q)^2-m^2+i\epsilon}\frac{[B(q)]^{k-1}}{(-i\lambda)^{k-2}} \text{.}
\end{equation}\\

\begin{figure}[h]
\begin{center}
\includegraphics[scale=0.5]{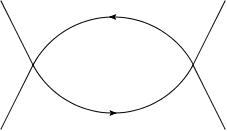} 
\caption{Fish-diagram.}
\label{fig:diagfish}
\end{center}
\end{figure}

In this expression,  $B(q)$ corresponds to the contribution of one bubble in the chain which is equivalent, of course, to the one-loop correction, s-channel, of the vertex, the so called   fish-diagram  (see Fig. \ref{fig:diagfish}). The factor $(-i\lambda)^{k-2}$ has been added to cancel the vertices that have been counted twice along the chain.\\

The expression for  $B(q)$ is a well known result \cite{bailin} 
\begin{equation}
B(q)=\frac{(-i\lambda)^2}{2}\int  \frac{d^4k}{(2 \pi)^4}\frac{i}{k^2-m^2+i\epsilon}\frac{i}{(k+q)^2-m^2+i\epsilon} \text{.}
\end{equation}

\begin{equation}\label{fishexact}
B(q)=\frac{-i\lambda^2}{32\pi ^2} \left[ \Delta - \int _0 ^1 dx \log \left( \frac{m^2 - q^2 x(1-x) - i\epsilon}{\mu ^2} \right) \right]  \text{,}
\end{equation}

\noindent where $\Delta$ is the divergent part that will cancel with the counterterms and $\mu$ is an arbitrary mass-scale  associated to the regularization procedure  which always appear when going through the renormalization program.\\

The contribution to the renormalon comes from the deep euclidean region in the integral \ref{Rk}, i.e. where  $-q^2\gg m^2$. In this way,  $B(q)$ in Eq. (\ref{fishexact}) can be approximated as:

\begin{equation}\label{fishDE}
B(q) \approx \frac{-i\lambda^2}{32\pi ^2} \log(-q^2/\mu^2) \text{,}
\end{equation}
\\

Inserting this result in Eq. (\ref{Rk}) and performing a Wick rotation, we find

\begin{eqnarray}
R_k(p)&=&\frac{-i\lambda^k}{(32\pi ^2)^{k-1}} \int \frac{d^4q}{(2 \pi)^4}\frac{1}{(p+q)^2+m^2} \nonumber \\
&\times& [\log(q^2/\mu^2)]^{k-1} \text{.}
\end{eqnarray}

This is an ultraviolet divergent expression. However, since the theory is renormalizable, we can separate this expression in a finite and a divergent part. We are only interested in the finite part. For this we expand the propagator in powers of $1/q^2$, neglecting the first two terms that leave divergent integrals. As we know this is justified due to the appearance of the usual counterterms.\\

So, we find
\begin{eqnarray}\label{R2}
R_k &=& -i \left( \frac{\lambda}{32\pi ^2} \right)^k 4 m^4 \int _\Lambda ^\infty dq [\log (q^2/\mu^2)]^{k-1}q^3 \nonumber \\
&\times&\left[\frac{m^4}{q^6}-\frac{m^6}{q^8}+...\right]\text{,} 
\end{eqnarray}
with $\Lambda >0$.
The dependence on the external momentum $p$ has disappeared. The lower limit $\Lambda $ comes form the fact that we are interested in the deep euclidean region, and has to be fixed in order to fulfill this condition.\\

Introducing $q=\mu e^t$ in the first two terms of Eq. (\ref{R2}), and fixing $\Lambda = \mu$, we find

\begin{eqnarray}
R_k &=&  -i \left( \frac{\lambda}{32\pi ^2} \right)^k  \frac{4 m^4}{\mu^2}\int_0 ^\infty dt e^{-2t} (2t)^{k-1} \nonumber \\
&\times& \left(1-\frac{m^2}{\mu^2} e^{-2t} \right)  \nonumber \\
&=&  -i \left( \frac{\lambda}{32\pi ^2} \right)^k \frac{2 m^4}{\mu^2} \Gamma (k) +i \left( \frac{\lambda}{64\pi ^2} \right)^k \frac{2 m^6}{\mu^4} \Gamma (k) \nonumber\\
&=& -i \left( \frac{\lambda}{32\pi ^2} \right)^k \widetilde{m}^2 \Gamma (k) +i \left( \frac{\lambda}{64\pi ^2} \right)^k \frac{m^2}{\mu^2}\widetilde{m}^2 \Gamma (k) \text{,} \nonumber \\
\end{eqnarray}
with $\widetilde{m}^2=2m^4/\mu^2$. We see that this diagram grows like $k!$, inducing then a pole in the Borel plane.\\

By taking the series $\Sigma R_k$,

\begin{equation}
D[\lambda] = \sum_k \frac{-i}{(32\pi ^2)^k}\Gamma(k)\lambda^k \text{.}
\end{equation}\\

its Borel transform is given by 
\begin{eqnarray}
B[b] &=& \sum_k \left( \frac{R_k}{\lambda^k} \right)\frac{b^{k-1}}{(k-1)!} \text{,}\nonumber \\
&=& -i \widetilde{m}^2\frac{1}{1-b/32\pi^2} +i \frac{m^2}{\mu^2}\widetilde{m}^2\frac{1}{1-b/64\pi^2} \text{,}
\end{eqnarray}

identifying, finally, the leading pole on the positive semi real axis in the Borel plane in $b=32\pi^2$ and the second pole in $b=64\pi^2$.\\
 
\section{Magnetic renormalons}\label{magneticrenormalons} 

We will now consider the renormalon-type diagram, taking magnetic effects into account.  The diagram to be calculated is the same one we considered previously, using now the weak field expansion of the Schwinger propagators. In this way we have

\begin{equation}\label{RBk}
R_{B,k}=\frac{1}{(-i\lambda)^{k-2}} \int \frac{d^4q}{(2\pi)^4} i D^B(p+q)[B_B(q)]^{k-1} \text{,}
\end{equation}

where  $D^B(p)$ is the weak expansion of the Schwinger propagator according to  Eq. (\ref{propmagneticodebil}). First let us consider one bubble 

\begin{equation}\label{fishmagnetico}
B_B(q) =  \frac{(-i\lambda)^2}{2} \int \frac{d^4k}{(2\pi)^4} i D^B(k)i D^B(q+k) \text{.}
\end{equation}

Splitting the propagator as

\begin{equation}
iD^{w}(k)=i\Delta_0(k) - i \Delta_B(k) \text{,}
\end{equation}

where the superindex $w$ signals the weak magnetic correction. $i\Delta_0$ is the vacuum propagator and $i\Delta_B$ includes the magnetic corrections up to order $|eB|^{2}$, according to  Eq. (\ref{propmagneticodebil})

In this way, we have now for Eq. (\ref{fishmagnetico}), 

\begin{eqnarray}
B_{w} (q)&& = \frac{(-i\lambda)^2}{2} \int \frac{d^4k}{(2\pi)^4} iD^{B,d}(k) iD^{B,d}(k+q) \nonumber \\
& &= \frac{(-i\lambda)^2}{2} \int \frac{d^4k}{(2\pi)^4} [i\Delta_0(k) - i \Delta_B(k)]\nonumber \\
&\times&[i\Delta_0(k+q) - i \Delta_B(k+q)] \text{.}
\end{eqnarray}

Since we want to include only terms up to order $|eB|^2$, neglecting higher powers of $|eB|$, we will deal with a vacuum contribution and two terms of order $|eB|^2$. These two terms are equal, as can be shown by means of a change of variable.  So we have

\begin{eqnarray}
B_{w}(q) &=& \frac{(-i\lambda)^2}{2} \int \frac{d^4k}{(2\pi)^4}
i\Delta_0 (k) \nonumber \\
&\times& [i\Delta_0(k+q) - 2i \Delta_B(k+q)] \text{,}
\end{eqnarray}
namely,

\begin{equation} \label{sumafish}
B_{w}(q)= \frac{(-i\lambda)^2}{2} \left\lbrace I_0 - 2(I_{1} + I_2) \right\rbrace \text{,}
\end{equation}
where,

\begin{equation}\label{formulaI0}
I_0 = \int \frac{d^4k}{(2\pi)^4} \frac{i}{k^2-m^2+i\epsilon}\frac{i}{(k+q)^2-m^2+i\epsilon}\text{,}
\end{equation}   

\begin{equation}\label{formulaI1}
I_1 = \int \frac{d^4k}{(2\pi)^4} \frac{i}{k^2-m^2+i\epsilon} \frac{i|eB|^2}{((k+q)^2-m^2+i\epsilon)^3} \text{,}
\end{equation} 

\begin{equation}\label{formulaI2}
I_2 = \int \frac{d^4k}{(2\pi)^4} \frac{i}{k^2-m^2+i\epsilon} \frac{2i|eB|^2 k_\perp^2}{((k+q)^2-m^2+i\epsilon)^4} \text{.}
\end{equation}

The first integral is the vacuum fish-diagram presented previously  (Eq. (\ref{fishDE})).

\begin{equation}\label{resI0}
I_0 \approx \frac{i}{16\pi ^2} \log(-q^2/\mu^2) \text{.}
\end{equation}

The problem then reduces to the calculation of the integrals $I_1$ and $I_2$ which can be handled using the standard Feynman parametrization together with a  Wick rotation. We find

\begin{eqnarray}
I_{1,E}&=& \frac{-i|eB|^2}{32\pi ^2}
\biggl[\frac{(2m^2 + q^2)}{m^2q^2(4m^2+q^2)}  \nonumber \\
&+& \frac{4m^2}{q^3(4m^2+q^2)^{3/2}}\log \left(  \frac{q-\sqrt{4m^2+q^2}}{q+\sqrt{4m^2+q^2}}  \right) \biggr]. 
\end{eqnarray}

The second integral $I_{2}$ can also be calculated in a standard way, getting

\begin{eqnarray}
&&I_{2,E} =  \frac{i|eB|^2}{48\pi ^2} 
\Bigg[  \frac{(3m^2 + q^2)}{m^2q^2(4m^2+q^2)} \nonumber \\
&+& \frac{6m^2+q^2}{2q^3(4m^2+q^2)^{3/2}}\log \left( \frac{q-\sqrt{4m^2+q^2}}{q+\sqrt{4m^2+q^2}}  \right) \nonumber   \\
&+& q_\perp ^2 \Biggl( \frac{-3(2m^2 + q^2)}{2q^4(4m^2+q^2)^2} \nonumber \\
&+& \frac{q^4+6m^4+4m^2q^2}{q^5(4m^2+q^2)^{5/2}} \log \left( \frac{\sqrt{4m^2+q^2}+q}{\sqrt{4m^2+q^2}-q}\right) \Biggr) 
\Bigg] \text{.} 
\end{eqnarray}

In the previous equations, the subindex E emphasizes that these integrals were calculated in the euclidean region. As we discussed in section \ref{secRenVacio}, the contribution to renormalons will come from the deep euclidean region ($q^2\gg m^2$). In this region we may approximate our results for integrals  $I_1$ and $I_2$, neglecting also in the denominators all terms with powers of $q$ higher than $2$, we get

\begin{equation}\label{resI1}
I_{1,DE}\approx \frac{-i|eB|^2}{32\pi ^2m^2q^2} \text{.}
\end{equation}\\

and

\begin{equation}\label{resI2}
I_{2,DE}\approx \frac{i|eB|^2}{48\pi ^2m^2q^2}.
\end{equation}\\

Summing according to Eq. (\ref{sumafish}), the different contributions given by  Eq. (\ref{resI0}), Eq. (\ref{resI1}) and Eq. (\ref{resI2}), the final expression for the magnetic-fish diagram in the weak field approximation is given by

\begin{equation}
B_{w}(q) \approx  \frac{-i\lambda ^2}{32\pi ^2}\left[\log(q^2/\mu^2) + \frac{|eB|^2}{3m^2q^2} \right].
\end{equation}

Now we have to insert this magnetic-fish term in the chain of bubbles that define the renormalon-type diagram. Note that in principle we could also have a contribution associated to the insertion of the vacuum chain of bubbles, without magnetic contribution, using the weak field expansion for the single remaining propagator in Fig.\ref{fig:diagrenormalon}. However, the last possibility does not produce a singularity in the Borel plane, because both contributions cancel among each other. See the appendix.  Therefore, we will concentrate on the chain of magnetic bubbles, of order $|eB|^{2}$, in the renormalon-type diagram of order $k$, summing then the series in $k$ and applying the Borel transform to see if we have  a Borel summable situation. We want also to compare with the previous existing analysis at finite temperature.

The relevant contribution is given by 

\begin{eqnarray}
R_{B,k}(p)&=&\frac{1}{(-i\lambda)^{k-2}} \int  \frac{d^4q}{(2\pi)^4} \Delta_0(p+q)
\left( \frac{-i\lambda ^2}{32\pi ^2} \right)^{k-1} \nonumber \\
&\times&\left[\log(q^2/\mu^2) + \frac{|eB|^2}{3m^2q^2} \right]^{k-1} \text{,}
\end{eqnarray}

and, as we noticed in section \ref{secRenVacio}, we will have a ultraviolet divergent integral. Nevertheless, since we are dealing with a renormalizable theory, we can expand the propagator $\Delta_0$ in powers of  $1/q^2$ and neglect the first two terms that give rise to divergent integrals. In the praxis, this corresponds to the inclusion  of the usual counterterms.

We have then to calculate

\begin{eqnarray}
R_{B,k}&=&\frac{1}{(-i\lambda)^{k-2}} \int \frac{d^4q}{(2\pi)^4}  \left[ \frac{m^4}{q^6}-\frac{m^6}{q^8}+ ... \right] \left( \frac{-i\lambda ^2}{32\pi ^2} \right)^{k-1} \nonumber \\
&\times& \left[\log(q^2/\mu^2) + \frac{|eB|^2}{3m^2q^2} \right]^{k-1}  \text{.}
\end{eqnarray}

Using the binomial theorem 

\begin{equation}
(A+B)^N=A^N+N\cdot A^{N-1}B + ... \quad  \text{,}
\end{equation}

we have

\begin{eqnarray}
&&R_{B,k}=-i \frac{\lambda^k}{(32 \pi^2)^{k-1}} \int  \frac{d^4q}{(2\pi)^4} \left[\frac{m^4}{q^6}-\frac{m^6}{q^8}+...\right] \nonumber \\
&\times&  \Biggl[\log(q^2/\mu^2)^{k-1} + \frac{(k-1)|eB|^2 (\log(q^2/\mu^2))^{k-2}}{3q^2m^2} + ... \Biggr] \text{.}  \nonumber \\ 
\end{eqnarray} 
Notice that the vacuum leading term as well as the next to leading order (NLO) vacuum term are can be extracted from the first two terms of the first square parenthesis together with the first term of the second square parenthesis. However, the leading term of the magnetic field is obtained by multiplying the first term of the first square parenthesis with the second term of the second square parenthesis. The next terms are subleading.

In this way, following the same procedure of section (\ref{secRenVacio}) and performing the angular integrals we find

\begin{eqnarray}
&&R_{B,k} = -i 4m^4 \left( \frac{\lambda}{32\pi ^2}  \right) ^k \int dq \Biggl[\frac{\left(\log (q^2/\mu^2)\right)^{k-1}}{q^3}  \nonumber \\
&-&\frac{m^2}{q^5}\log (q^2/\mu^2)^{k-1}+ \frac{(k-1)|eB|^2 (\log(q^2/\mu^2))^{k-2}}{3q^5m^2} + ... \Biggr]  \text{.} \nonumber \\
\end{eqnarray}

Using the change of variable  $q=\mu e^t$, $dq=\mu e^tdt$, 

\begin{eqnarray}
&&R_{B,k} = -i \left( \frac{\lambda}{32\pi ^2} \right)^k \frac{4m^4}{\mu^2} \int dt \Biggl[e^{-2t}(2t)^{k-1} \nonumber \\
&-& \frac{m^2}{\mu^2}e^{-4t} (2t)^{k-1} + \frac{(k-1)|eB|^2 e^{-4t}(2t)^{k-2}}{3m^2 \mu^2} + ... \Biggr]  \text{,} \nonumber \\
\end{eqnarray}
we see the appearance of the Gamma function in both terms, and using the definition of $\widetilde{m}$, we have 

\begin{eqnarray} \label{rrr}
R_{B,k} &=& -i \widetilde{m}^2 \left( \frac{\lambda}{32\pi ^2}  \right)^k \Gamma(k) +i \left( \frac{\lambda}{64\pi ^2} \right)^k \frac{m^2}{\mu^2}\widetilde{m}^2 \Gamma (k) \nonumber \\
 &-&2i \frac{\widetilde{m}^2}{\mu^2} \left( \frac{\lambda}{64\pi ^2}  \right) ^k \frac{|eB|^2}{3m^2} \Gamma(k) + ... \hspace{1mm}  \text{,}
\end{eqnarray}

where we have used also the property $\Gamma(z+1)=z\Gamma(z)$.

Now we can find the Borel transform $B[b]$ of $\Sigma R_{B,k}$,

\begin{eqnarray}
B[b]&=&\sum_k \left( \frac{R_{B,k}}{\lambda ^k}  \right) \frac{b^{k-1}}{(k-1)!} \text{,}\nonumber \\
& =& - i\widetilde{m}^2 \sum_k \left( \frac{1}{32\pi ^2}  \right) ^k b^{k-1} \nonumber \\
&+&\left(\frac{i m^2}{\mu^2}\widetilde{m}^2-\frac{2i|eB|^2 \widetilde{m}^2}{3m^2 \mu^2 }\right) \sum_k \left( \frac{1}{64\pi ^2}  \right)^k b^{k-1} + ... \quad \text{.} \nonumber \\
\end{eqnarray}

These sums correspond to well known geometrical series, obtaining then our final result

\begin{eqnarray}
B[b] &=& \frac{-i \widetilde{m}^2}{b-32\pi ^2} \nonumber \\
&+&\left(i \frac{m^2}{\mu^2}\widetilde{m}^2- \frac{2i|eB|^2 \widetilde{m}^2}{3m^2 \mu^2}\right) \frac{1}{b-64\pi ^2} + ...  \text{.} 
\end{eqnarray}

\begin{figure}[h]
\begin{center}
\includegraphics[scale=0.5]{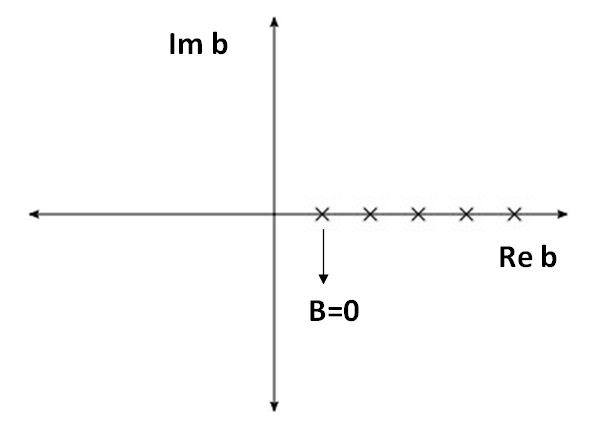}
\caption{Poles in the Borel plane. These are located on the semi-positive real axis in  $b=n \cdot 32\pi ^2$, with $n$ a natural number.}
\label{fig:polos}
\end{center}
\end{figure}

This expression shows the appearance of a new subleading  pole in the Borel plane associated to magnetic effects, together with the NLO vacuum pole. The magnetic contribution is quite small in comparison with the vacuum NLO term. From general principles \cite{parisi,david}, we know that the location of these poles will be always a  multiple of the first coefficient of the beta function. However, the first coefficient of the beta function in $\lambda \phi^4$ theory is given by $\beta_0= 3/16\pi^2$.  In the present calculation we are dealing only with one channel and the correct beta function comes from taking into account the contributions from the  s, t and u channels, which explain the discrepancy  with the factor 3. Therefore we obtain $b= n \times 2/(3 \beta_0)$ instead of $b= n \times 2/\beta_0$ with n a natural number. In Fig. \ref{fig:polos} a scheme of the location of these poles is shown.

Although the location of the poles does not depend on the field intensity, the residues of these poles do depend on it. As we expected, the evolution of the residues is opposite with respect to the thermal evolution found previously in the literature \cite{Loewe},

\begin{eqnarray}\label{tempe}   
R_{T,k} \propto -i \left(\frac{\lambda}{32 \pi^2} \right)^k \Gamma(k) -2 i f_2(1/T) \left(\frac{\lambda}{64 \pi^2} \right)^k \Gamma(k), \nonumber \\ 
\end{eqnarray}

where 

\begin{equation}
f_2(1/T) \approx -4 \int_0^\infty \frac{dx x^{1/2}}{\sqrt{x+1}}e^{-(1/T) \sqrt{x+1}} < 0.
\end{equation}

In order to make a better comparison between Eq. (\ref{rrr}) and Eq. (\ref{tempe}), we will write  Eq. (\ref{rrr})  as follows

\begin{eqnarray}   
R_{B,k} &\propto& -i \left(\frac{\lambda}{32 \pi^2} \right)^k \Gamma(k) +i \left( \frac{\lambda}{64\pi ^2} \right)^k \frac{m^2}{\mu^2}\widetilde{m}^2 \Gamma (k)\nonumber \\
&-&2 i f_1(qB) \left(\frac{\lambda}{64 \pi^2} \right)^k \Gamma(k),  
\end{eqnarray}
with $f_1(qB)=\frac{\widetilde{m}^2|eB|^2}{\mu^2 3m^2}$. We plot these functions separately $f_1$ y $f_2$ in Fig. 4 and  Fig. 5 respectively.

\begin{figure}[h]
\begin{center}
\includegraphics[scale=0.28]{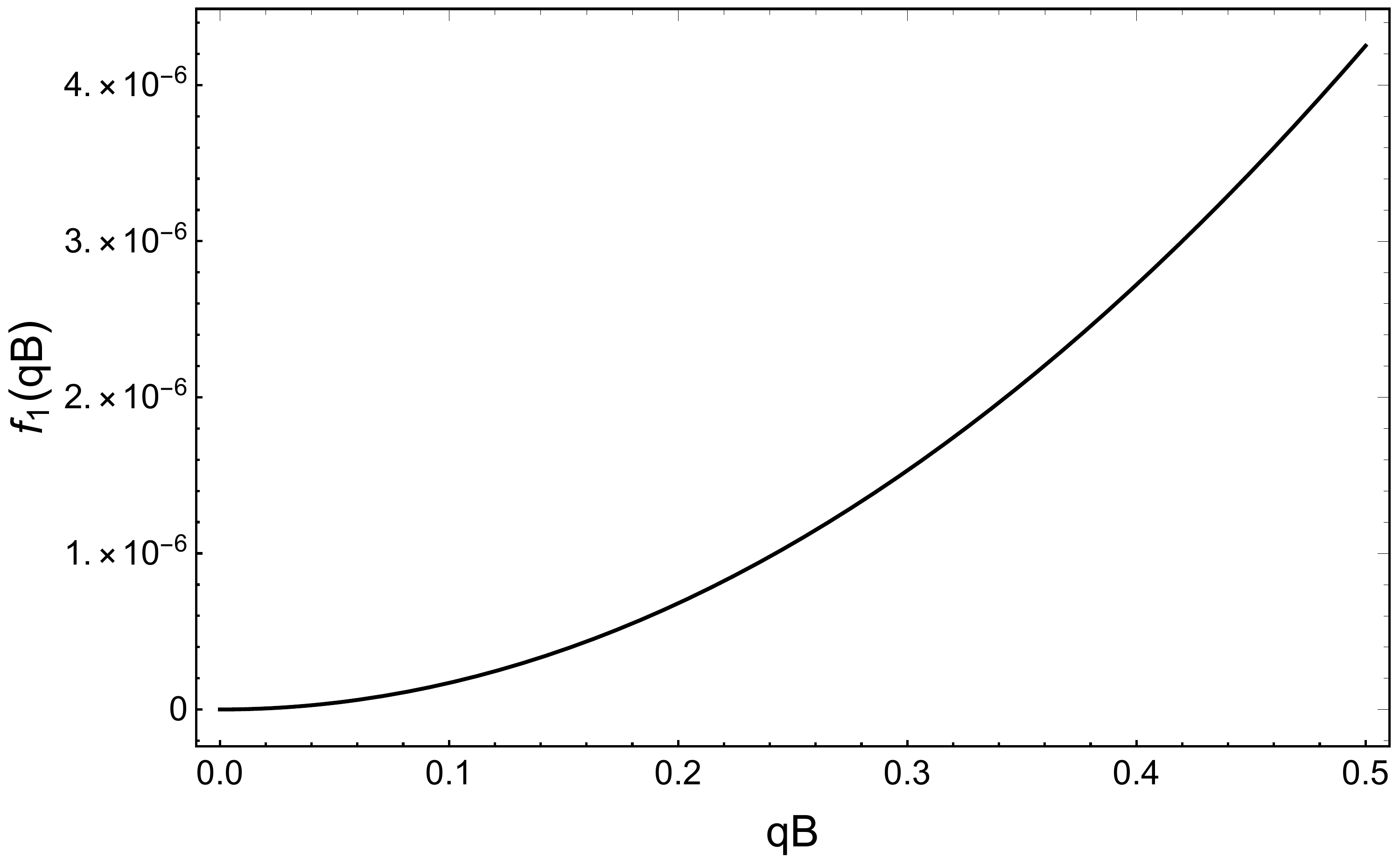}
\caption{$f_1(qB)$vs. $qB$}
\label{fig:grafresiduos2}
\end{center}
\end{figure} 

\begin{figure}[h]
\begin{center}
\includegraphics[scale=0.28]{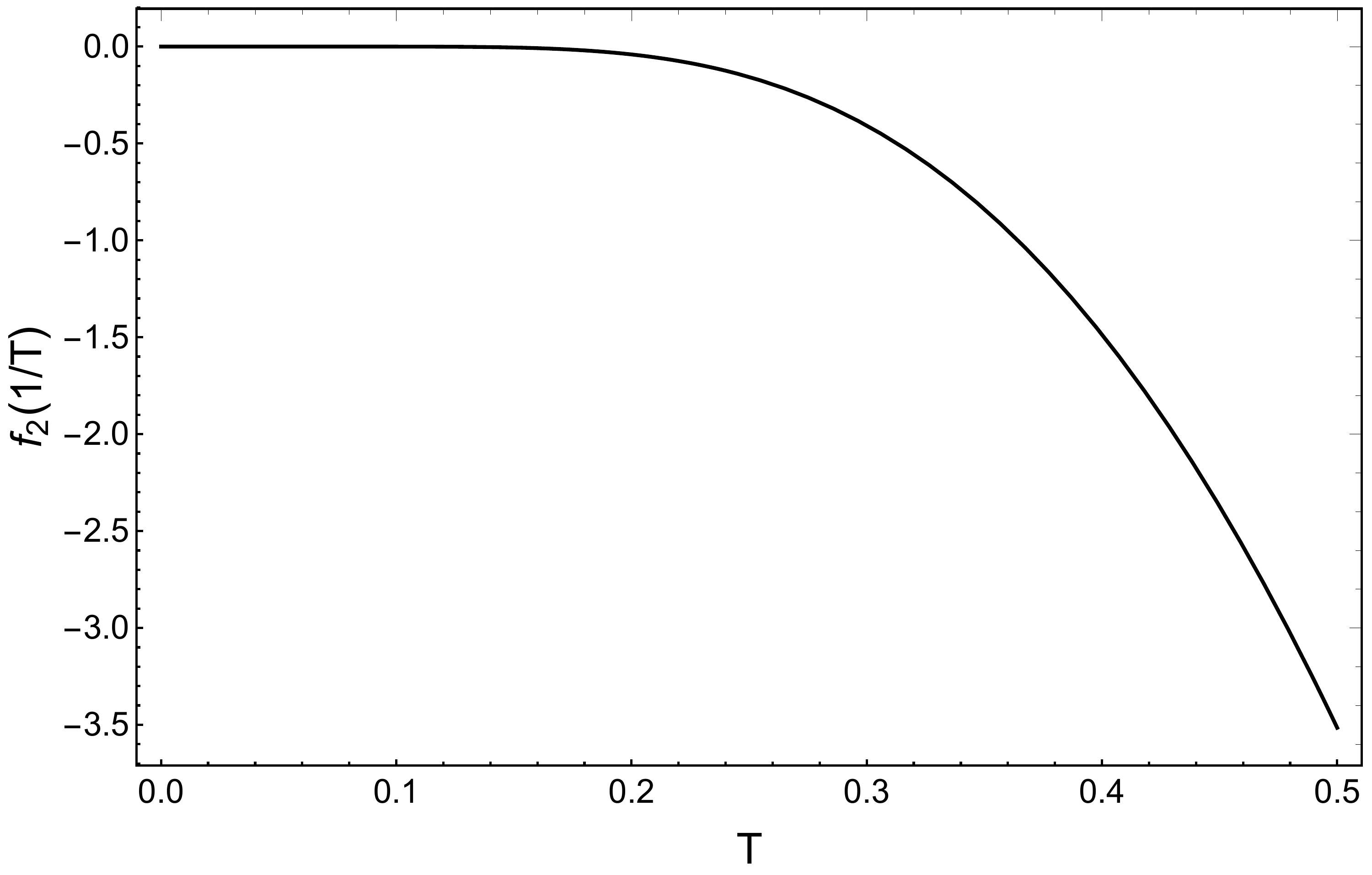}
\caption{$f_2(1/T)$ vs. $T$}
\label{fig:grafresiduos1}
\end{center}
\end{figure}

It is important to stress that in region of a very intense magnetic strength, which is handled in terms of the lowest Landau level approximation (LLLA) for the Schwinger propagator, we do not find the presence of new magnetic renormalons. It is possible to write the Schwinger propagator (excluding the phase) as a sum over Landau levels according to
\begin{equation}
iD^{B} = 2 \sum _{0}^{\infty}(-1)^{l} L_{l}\left(\frac{2k_{\perp}^{2}}{|qB|}\right ) e^{- k_{\perp}^{2}/|qB|}i\Delta_{l}^{B}(k _{\parallel}), 
\end{equation}

\noindent where we have introduced the ``parallel" propagators
\begin{equation}
i\Delta ^{B}_{l} (k_{\parallel}) = \frac{i}{k_{\parallel}^{2} - (2l +1)qB - m^{2} + i\epsilon}.
\end{equation}

\noindent When $|eB| \gg m^{2}$ (strong magnetic field region) and $|eB|$ bigger than the momentum, the gap between the different energy levels, given by $E_{l} = \sqrt{ (k_{3})^{2} + (2l +1)|eB| + m^{2}}$  also grows. If there are no additional external sources that may excite higher energy levels, a good approximation is to take  only the lowest Landau level in the general expression given above. In this way, when $|eB| \rightarrow \infty$  we have

\begin{equation}
i D_{B}(k)  \approx 2i \frac{e^{-k_{\perp}^{2}/|eB|}}{k_{\parallel}^{2} - |eB| - m^{2}}
\end{equation}

\noindent Notice that in this approximation the perpendicular and longitudinal contributions decouple from each other. If we calculate the renormalon-type diagrams of order k, using this approximation for the magnetic propagators, we get a term proportional to $k!/ k^{k}$. This produces a convergent series which is, of course,  Borel summable.

\section{Conclusions} \label{conclusions}
We have analyzed the effects of an external constant magnetic field on a typical renormalon-type diagram in the  $\lambda \phi ^{4}$ theory, using a weak field approximation for the Schwinger propagator. In fact, we discovered the existence of poles in the Borel plane, subleading with respect to the vacuum usual singularity, which depend on the strength of the magnetic field. 

An interesting observation, if we compare our result with a previous analysis of this problem at finite temperature, is that now, the residues grow with the magnetic field, whereas the opposite effect was reported in the thermal case. In principle we may understand such behavior in terms of a simple phenomenological argument. In general the residue of the propagator is absorbed in the wave function renormalization. So, we could think that due to structure of Landau levels that emerge, the probability of interaction increases. The opposite effect is induced by thermal corrections. In other words, the magnetic moments associated to the Landau ``orbits" tend to stay aligned due to the influence of the external magnetic field, whereas temperature induces a disorder effect. It should be noted that the B-dependent NLO renormalons are largely screened and very negligible as compared with the vacuum NLO renormalon.

There are several physical parameters where the thermal and magnetic effects are opposite to each other. Recently, for example,  this behavior has been reported  for $\pi$-$\pi$ scattering lengths \cite{pi1,pi2}. In the context of QCD sum rules, it has been found that the continuum threshold grows as function of the magnetic field whereas it diminishes with temperature. The same, for example, occurs with the behavior of the pion decay constant $f_{\pi }$ \cite{sum}.

\section*{ACKNOWLEDGMENTS}

M. Loewe acknowledges support from FONDECYT (Chile) under grants No. 1170107, No. 1150471, No. 1150847 and ConicytPIA/BASAL (Chile) grant No. FB0821, M. Correa and D. Valenzuela acknowledge support from FONDECYT (Chile) under grant No. 1170107 and R. Zamora would like to thank support from CONICYT FONDECYT Iniciaci\'on under grant No. 11160234.

\section*{Appendix}

In this appendix we will calculate the diagram shown in Fig. {\ref{fig:diagrenormalon}}, by taking the magnetic correction only in the single scalar propagator without considering magnetic corrections in the chain of bubbles, after a Wick rotation we have

\begin{equation}\label{RBk}
R_{B,k}=\frac{1}{(-i\lambda)^{k-2}} \int \frac{d^4q}{(2\pi)^4} D^B(p+q)[B_B(q)]^{k-1} \text{.}
\end{equation}

We use the Schwinger's propagator in Euclidean space in the weak field expansion

\begin{eqnarray}\label{propmagneticodebil2}
iD^B(k)&&\xrightarrow{eB\rightarrow 0} \frac{-i}{k_\parallel^2 + k_\perp^2+ m^2} + \frac{i(eB)^2}{(k_\parallel^2 + k_\perp^2 +m^2)^3} \nonumber \\
&-& \frac{2i(eB)^2k_\perp^2}{(k_\parallel^2 + k_\perp^2 +m^2)^4} \text{,}
\end{eqnarray}

and 

\begin{equation}\label{fishDE2}
B(q) \approx \frac{-i\lambda^2}{32\pi ^2} \log(q^2/\mu^2) \text{.}
\end{equation}

As we are interested only in the calculation of the magnetic contribution, we will calculate the second and third terms of

\begin{eqnarray}
\text{I} &=&\frac{i \lambda^k}{(32 \pi^2)^{k-1}} (eB)^2 \int \frac{d^4q}{(2\pi)^4} \frac{1}{((p+q)^2+m^2)^3}\nonumber \\
&\times&[\log(q^2/\mu^2)]^{k-1}. 
\end{eqnarray}

Expanding in powers of $1/q^2$

\begin{eqnarray}
\text{I} =i\left(\frac{\lambda}{32 \pi^2}\right)^{k} 4 (eB)^2 \int_\Lambda^\infty dq \frac{1}{q^3}[\log(q^2/\mu^2)]^{k-1}, \nonumber \\
\end{eqnarray}

and using the change of variable $q=\mu e^{t}$

\begin{eqnarray}
\text{I} =i\left(\frac{\lambda}{32 \pi^2}\right)^{k} 4 \frac{(eB)^2}{\mu^2} \int_0^\infty dq e^{-2t}(2t)^{k-1}.
\end{eqnarray}

Now we will calculate II

\begin{eqnarray}
\text{II}  &=&\frac{-i \lambda^k}{(32 \pi^2)^{k-1}} 2(eB)^2 \int \frac{d^4q}{(2\pi)^4} \frac{(p+q)^2_{\bot}}{((p+q)^2+m^2)^4}\nonumber \\
&\times&[\log(q^2/\mu^2)]^{k-1},
\end{eqnarray}

we expand in powers of $1/q^2$, note that $q_\bot=q \sin\theta \sin\chi$ (polar coordinates in a 4-dimensional Euclidean space)

\begin{eqnarray}
\text{II}&=&-i\frac{\lambda^k}{(32\pi^2)^{k-1}}  \frac{2(eB)^2}{(2 \pi)^4} \int_\Lambda^\infty dq \int_0^{2\pi} d\phi \int_0^\pi d\theta \sin^2\theta   \nonumber \\ 
&\times& \int_0^\pi d\chi  \sin\chi \frac{ \sin^2\theta \sin^2\chi}{q^3}[\log(q^2/\mu^2)]^{k-1},
\end{eqnarray}

\begin{eqnarray}
\text{II} =-i\left(\frac{\lambda}{32 \pi^2}\right)^{k} 4 (eB)^2 \int_\Lambda^\infty dq \frac{1}{q^3}[\log(q^2/\mu^2)]^{k-1}, \nonumber \\
\end{eqnarray}

using the change of variable $q=\mu e^{t}$

\begin{eqnarray}
\text{II} &=&-i\left(\frac{\lambda}{32 \pi^2}\right)^{k} 4 \frac{(eB)^2}{\mu^2} \int_0^\infty dq e^{-2t}(2t)^{k-1} \nonumber \\
 &=& \text{-I},
\end{eqnarray}

therefore, these terms cancel among each other.



\end{document}